\documentclass[12pt,twoside,notitlepage]{report}

\input BoxedEPS
\SetRokickiEPSFSpecial  
\HideDisplacementBoxes

\def\thruD#1{\mathrel{\mathop{#1\!\!\!\!/}}}
\let\cal\mathcal

\begin{document}
\raggedbottom
\clubpenalty=15000
\widowpenalty=15000

\renewcommand{\topfraction}{0.75}
\renewcommand{\bottomfraction}{0.75}
\renewcommand{\textfraction}{0.1}

\title{One Theorist's Perspective on Four Eras\\
 of   Electron-Proton Scattering}
\author{Robert L. Jaffe}
\date{\small\it Center for Theoretical Physics,\\ 
Laboratory for Nuclear Science, and
Department of Physics \\
Massachusetts Institute of Technology,
Cambridge, Massachusetts 02139\\[0.5ex]
{\footnotesize\rm
MIT-CTP-2791\qquad hep-ph/9811327}\\
\rm A transcription with minor alterations of the
talk presented\\[-0.5ex]
to honor Sid Drell on the occasion of his retirement.}
\maketitle

\setcounter{page}{1}

\def\kp{\kern0.05em}
\noindent
{\Large I}{\footnotesize\uppercase{ wou\kp ld l\kp i\kp k\kp e}} to thank the organizers
for the invitation to speak on  this great occasion.  
To set the tone, I'd like to begin with a couple of quotes from our 
honored guest.  The first comes from the 1973 Photon-Lepton Conference 
held in Bonn.  Sid introduced his talk on Deep Inelastic Physics with
\begin{quote}\small\baselineskip=0.9\baselineskip
  Ancient explorers in search of distant lands and 
	treasures had no idea how great the challenge or how difficult 
	the passage.  They sailed into uncharted seas.  Their only 
	scale of distances came from previous journeys starting from 
	ancient Phoenicia and \ldots Crete, through the Mediterranean 
	to the North Sea.
	
         \ldots We can look back today with ancient mariners
and view 
	how far we have progressed \ldots in our explorations of the 
	Lepton [Hadron] frontiers of Nature~\cite{1}.
\end{quote}
I have to confess: at the time, I had little idea what Sid was talking 
about.  Now I understand that he was planting in the historical record
a classical context for the talk I am about to give.

As all of you know, Sid  was a much more practical man than 
this quote would suggest.  His down-to-earth attitude toward 
particle theory rubbed off on most of his students.  It is exemplified by the 
following quote from an afternoon discussion section at the 1969 Erice 
School.  Sid was asked what he thought of the attempt to understand 
deep inelastic phenomena with the aid of the DGS representation.  You 
don't have to know anything about the DGS representation to appreciate 
his response,\footnote{Though hearing it in Sid's Atlantic City accent 
helps.}
\begin{quote}\small\baselineskip=0.9\baselineskip
	    I just want to make a remark about the use of the 
	    DGS representation.  When I started this problem, I 
	    studied this representation in detail and found that it 
	    was, to me, void of any physics.  Therefore I put this DGS 
	    representation aside and I will not use it again, because 
	    with pure mathematical jiggling around you can't solve 
	    physics.  I can't understand how to even approximately 
	    introduce physics into the DGS spectral function and so I 
	    give it up~\cite{2}.
\end{quote}
This practical, explicit, straightforward attitude toward theoretical 
physics permeated Sid's teaching and was inherited by all the people 
who worked with him.\footnote{I apologize to D (Deser), G (Gilbert), 
and S (Sudarshan)\ldots. It certainly wasn't their fault, and they 
have made many great contributions to physics and biology.}
  
When Mike Peskin called up and asked me to talk about ``Four Eras in 
Electron-Proton Scattering'', I was too ashamed to tell him I didn't 
know what the four eras were.  So, I decided I would have to make them 
up myself.  I make no claims of historical accuracy, nor is this a 
review.  Instead, borrowing the image from Feynman's 
sum-over-histories, you should view this talk as one particular path 
in the path integral.  I'm not sure my contribution to the 
sum-over-paths is very large.  I know I will be idiosyncratic, I hope 
to be provocative.  With this in mind, I have changed Mike's suggested 
title slightly, to ``One Theorist's Perspective on Four Eras of 
Electron-Proton Scattering''.

Inspired by Mike's suggested title and Sid's first quote, I have 
chosen to organize my talk in a somewhat classical form.

In the beginning was an ``Archaic Era" (1954--1966), when certain 
primitive cultures flourished, distinguished by an unusual rite known 
as ``Electron Scattering''.  One developed on the shores of San 
Francisco Bay.  Remains of other contemporary and similar cultures can 
be found in Cambridge, Massachusetts and in Hamburg, Germany.  During 
the late 1960s the primitive culture by the Bay experienced a 
``Classical Era" (1966--1972) when great cultural archetypes were 
crafted.  There was a rich interaction between the ``priestly'' (read 
``theoretical'') and ``artisan'' (read ``experimental'') classes.  
Like many great cultural awakenings, it was catalyzed by an invasion
of  barbarians, who challenged the established order and enriched the
gene  pool~\cite{3}.

All of this climaxed in the emergence of QCD in 1970--1972.  Then 
followed a ``Hellenistic Era" (1972--1980) when electron scattering 
cultures spread throughout the world.  Many migrated away from the 
center of high culture in California, and the practice of Deep Inelastic 
Scattering became ritualized.  The local culture here at SLAC turned 
away from its classic roots toward different goals.  One might call 
this a ``Dark Age'', but given the accomplishments of the intervening 
years, I don't think it's appropriate.  Finally,  a 
Renaissance in deep inelastic physics arose in the 1980s and continues 
today.  The classic rituals were rediscovered and reinterpreted as 
probes of the parts of QCD that we really don't understand.

\begin{figure}[htb]
$$
\BoxedEPSF{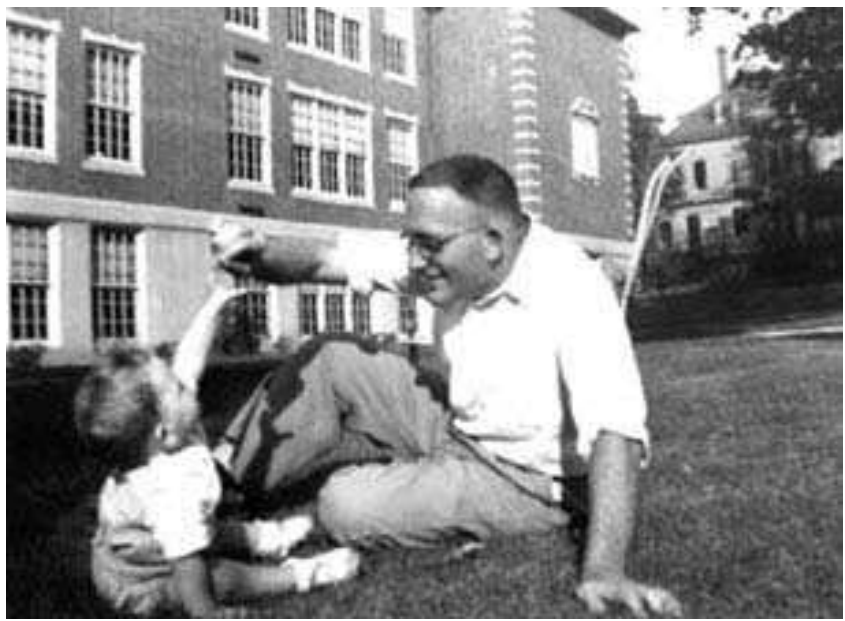 scaled 1600}  
$$
\caption{Sid and his son Daniel, around 1954.}
\label{Sid1}
\end{figure}

As a way of keeping time as I tell this story, I managed, with 
Harriet's help, to find pictures of Sid from each of the four eras.  
So our clock begins with Figure~\ref{Sid1}, 
a picture of Sid in Boston 
with his son, Daniel, before his migration to California.  The year 
was 1954, when the first elastic electron-proton scattering 
experiments were undertaken by Robert Hofstadter and collaborators at 
Stanford.

\section*{The Archaic Era, 1954--1966}

\begin{figure}[hbt]
$$
\BoxedEPSF{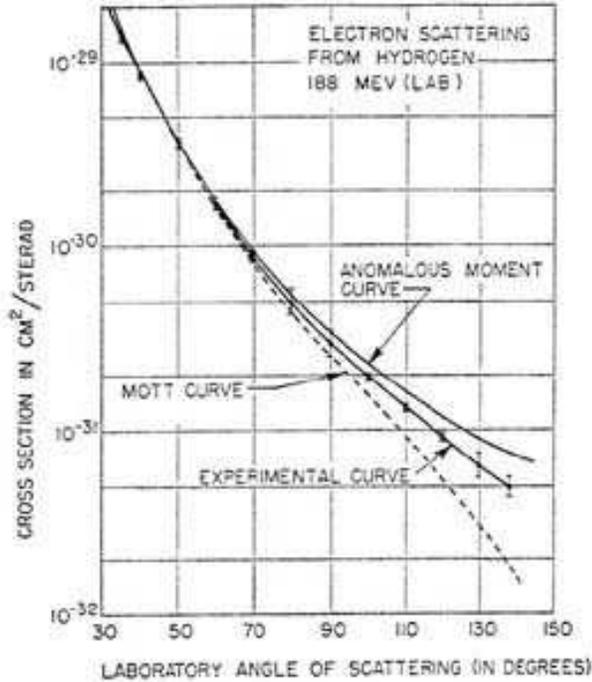 scaled 1800}  
$$
\caption{First data on the proton form factor from McAllister and Hofstadter, 1955.}
\label{MacHof}
\end{figure}

Hofstadter's experiments were incredibly important and influential.  
The form factors of the proton were measured for the first time, and 
showed the nucleon to be composite.  Figure \ref{MacHof}, from 
McAllister and Hofstadter's first paper on the subject~\cite{4}, shows that
the  experimental cross-section could not be fit if the proton was treated 
as a pointlike Dirac particle (the ``Mott'' cross section), nor with 
the addition of an anomalous magnetic moment alone.  More structure
was  required.

Sid became very involved in this physics.  Sid's early papers, to 
which Mike has already referred, explored the physics behind the form 
factors of the proton, the neutron, and nuclei.  In fact, in a very 
real sense, Sid wrote the book on the subject:  Drell and 
Zachariasen's {\sl Electromagnetic Structure of Nucleons}, 
published in 1961,	 is the classic monograph
on the subject~\cite{5}.

Already at this time it was becoming clear that nucleons are {\it 
very\/} composite.  Calculations in model field theories in which a 
pointlike nucleon interacts with pointlike pions could account for the 
nucleon's magnetic moment and charge radius, which parameterize the 
form factor near zero momentum transfer, but they could not account 
for the observed rapid falloff of the form factor at momentum 
transfers of order 1 GeV. The form factor fell like a dipole, ${\cal 
O}(1/Q^{4})$ rather than a monopole, as predicted by 
``vector dominance''~\cite{6}. Protons seemed to be very squishy 
indeed.  

Sid pioneered the use of dispersion relations to characterize form 
factors.  When dispersed in the current channel, the form factor is 
expressed in terms of mesons with $1^{--}$ quantum numbers, an 
approach that led to vector dominance.  When dispersed in the nucleon 
leg, an approach pioneered by Sid and his wonderful  student Heinz 
Pagels~\cite{7}, the 
nucleon's compositeness is related to the baryon resonances with the 
same quantum numbers as the nucleon.  In either case, the 
compositeness of the nucleon was related to the proliferation of 
hadrons that was occurring in the late 1950s and early 1960s.  
Compositeness has thoroughly dominated the thinking about hadron 
structure ever since.

During the later Archaic period, physicists started to ask how to 
obtain more detailed information about the structure of nucleons.  A 
few people started thinking about {\it inelastic\/} electron 
scattering at that time; Sid was one of the first.  Mostly, however, 
the community focused on elastic scattering.  In 1962 Carl Barber, 
then a collaborator of Hofstadter's at Stanford, wrote a review on 
inelastic electron scattering from nuclei~\cite{8}. Apparently there 
was so little interest in the potential of inelastic scattering to 
elucidate the structure of the nucleon that Barber specifically 
excluded particle production, that is, pion production, from his 
30-page review.

The proliferation of hadronic resonances and the apparent 
compositeness of the nucleon led to a picture of the strong 
interactions that is now very difficult to appreciate.  The idea was 
called ``nuclear democracy'', or the ``bootstrap''.  It was a deep and 
subtle picture, founded on dispersion theory, and like many stations 
along the way to QCD, it contains much that remains true.  However, it 
gave little help to young students, like myself at the time, who 
sought to visualize the nucleon.  In those days when one asked the 
question, ``What is a proton?''  one was told, ``It's a neutron plus a 
pion.''  When one asked ``What is a neutron?''  one was told, ``It's a 
proton plus a pion.''  And when one asked, ``What's a pion?''  one was 
told, ``It's a proton plus an antineutron.''  It is hard to 
appreciate, in retrospect, how seriously this was taken and how 
passionately it was studied.  Inelastic electron scattering did not 
seem like a promising way to explore the bootstrap.

Of course, continuum inelastic electron scattering was first carried 
out in the famous MIT-SLAC program headed by Friedman, Kendall, and 
Taylor.  Those early experiments are the topic of books, 
so I won't really cover them in any detail~\cite{9}.
%
%
Jerry Friedman tells me that the principal goals of the experiment 
were first elastic scattering, then electroproduction of resonances, and 
finally continuum electroproduction, for which they had few 
expectations.  Sid, however,  took inelastic electron scattering
very  seriously.  In 1964 he and Dirk Walecka published an {\it Annals of
Physics\/}  article that, drawing upon earlier work by Bjorken, von~Gehlen,
Gourdin, and Hand~\cite{10}, defined the inelastic kinematics and the famous 
structure function $W_{\mu\nu}$~\cite{12}. It became a bible for those 
of us who wanted to study the subject.

Now I have arrived at 1966, when things really began to change.  An 
excellent perspective on the state of particle theory can be found by 
looking at the proceedings of the Berkeley Conference, which was the 
Rochester meeting of that year.  In the introductory session, Murray 
Gell-Mann talked about quarks.  His remarks give one a glimpse at the
dominance of nuclear democracy and the dilemma facing anyone who 
thought seriously about quarks at that time.  Quoting from 
Murray~\cite{13},
\pagebreak[2]
\begin{quote}\small\baselineskip=0.9\baselineskip
	We consider three hypothetical and probably fictitious 
	spin-1/2 quarks\ldots.
	
	Now what is going on?  What are these quarks?  It is 
	possible that real quarks exist, but if so they have a high 
	threshold for copious production, many [G]eV; if this 
	threshold comes from their rest mass, they must be very heavy 
	and it is hard to see how deeply bound states of such heavy 
	real quarks could look like $q\bar q$, say, rather than a 
	terrible mixture of $q\bar q$, $qq\bar q\bar q$, and so on.  
	Even if there are light real quarks, and the threshold comes 
	from a very high barrier, {\it the idea that mesons and 
	baryons are made primarily of quarks is difficult to believe, 
	since we know that, in the sense of dispersion theory, they 
	are mostly, if not entirely, made up out of one another\/} [my italics]. 
\end{quote}
 So this is the bootstrap philosophy.  Murray goes on,
\begin{quote}\small\baselineskip=0.9\baselineskip
	The probability that a meson consists of a real quark pair 
	rather than two mesons or a baryon and antibaryon must be 
	quite small.  Thus it seems to me that whether or not real 
	quarks exist, the $q$ and $\bar q$ we have been talking about are 
	mathematical; in particular, I would guess that they are 
	mathematical entities that arise when we construct 
	representations of current algebra\ldots.  Their effective masses, to 
	the extent that these have meaning, seem to be of the order of 
	one-third the nucleon mass.  One may think of mathematical quarks as 
	the limit of real light quarks confined by a barrier, as the barrier 
	goes to an infinitely high one\ldots.
	
	If the mesons and baryons are made of mathematical quarks, 
	then the quark model may perfectly well be compatible with the 
	bootstrap hypothesis, that hadrons are made up out of one 
	another.
\end{quote}
Here, clearly, is a great physicist struggling with the apparent 
contradiction that dispersion theory tells us that hadrons are 
composed on one another, and spectroscopy had begun to suggest that 
hadrons are made up of quarks.  It was a time of great 
dissonance.\footnote{Bj reminds me that the conflict 
between dispersion theory and the quark model still has not been laid 
to rest.  For some thoughts on the subject, see Refs.~\cite{14,15}.}

Later in the Berkeley Conference, in his very last remarks as rapporteur of the 
electromagnetic interactions session, Sid said something remarkable, 
which pointed the way to the future.  After delivering his talk, and 
after answering all questions from the audience, Sid 
asked himself a question~\cite{16},
\begin{quote}\small\baselineskip=0.9\baselineskip
	What would I like to see measured?  Let me just say briefly 
	that I'd much like to see inelastic electron or muon cross 
	sections measured; they provide the inelastic nucleon form factors 
	that are of great interest in their own right.  Moreover they are 
	also the necessary input that goes into the neutron-proton 
	mass-difference calculations, if their isovector structure can be 
	measured, or into the hyperfine-structure calculation, if their spin 
	structure can be measured. 
\end{quote}
And then,
\begin{quote}\small\baselineskip=0.9\baselineskip
	{\it Also there are some sum rules, asymptotic statements derived 
	by Bjorken and others, as to how these inelastic cross 
	sections should behave in energy\/} \hbox{[my~italics].} 
\end{quote}
This, of course, was a prescient remark, made very 
early in the history of the subject.  I want to spend a couple of 
transparencies discussing it.

This thread goes back to the work of Ken Johnson, who in 1961 wrote a 
little-recognized paper on the relation between the amplitudes that 
are measured in current-hadron scattering and the time-ordered 
products ($T$-products) of local operators~\cite{17}. Johnson pointed 
out that the relationship is not as direct as is taught in field theory 
courses.  Symmetries can require and infinities can generate extra 
terms that must be removed from the product of local operators before 
they can be related to physical amplitudes.  He called his new objects 
$T^{\star}$-products.  They differ from $T$-products by local terms 
that generate polynomials functions of the momenta in scattering 
amplitudes.  It doesn't sound like much.  In 1966, Johnson and 
Low~\cite{18} and Bjorken~\cite{19} independently explored how to 
extract $T$-products from scattering amplitudes.  They introduced a 
limit (the Bjorken-Johnson-Low limit), where the energy transfer
($q^{0}$) is taken to imaginary  infinity to isolate operator products, 
and then explored the implications.  They wrote two very different and 
both very influential papers.  Johnson and Low explored what happened 
in perturbation theory.  They discovered anomalies that were 
precursors of anomalous dimensions that now figure so centrally in 
deep inelastic phenomenology.  Bjorken, the optimist, examined the 
implications of the BJL limiting procedure under the hypothesis that 
{\it products of operators have free field singularities at zero 
separation}.  Bjorken's paper, with the formidable title ``Applications of the
Chiral $U(6)\times U(6)$ Algebra of Current  Densities'', founded the field
of deep inelastic physics.  He took the 
$q^{0}\to i\infty$ limit, which we now understand to be the deep 
inelastic limit, and when he encountered products of local operators, 
he replaced them by their values in free field theory,
\begin{equation}
    \Bigl[J_{\mu}^{a}(\vec x),J_{\nu}^{b}(\vec y)\Bigr] \to \cdots \,
    \delta^{3}(\vec x - \vec y) \theta_{\mu\nu}^{ab}(\vec x)\ .
    \label{eq:1}
\end{equation}
At the time Bjorken was cautious, but the implication is that there 
are pointlike constituents inside hadrons, whether you say it 
explicitly or not.  Bjorken's assumption of free-field behavior at 
short distances  was quite radical, and Johnson and Low had already 
shown that it is violated in perturbation theory.  Now we know that 
it's almost correct: the corrections are only logarithmic.

To capture the spirit of the times, I have reproduced a small section 
of Bjorken's paper. 
\begin{figure}[htb]
$$
\BoxedEPSF{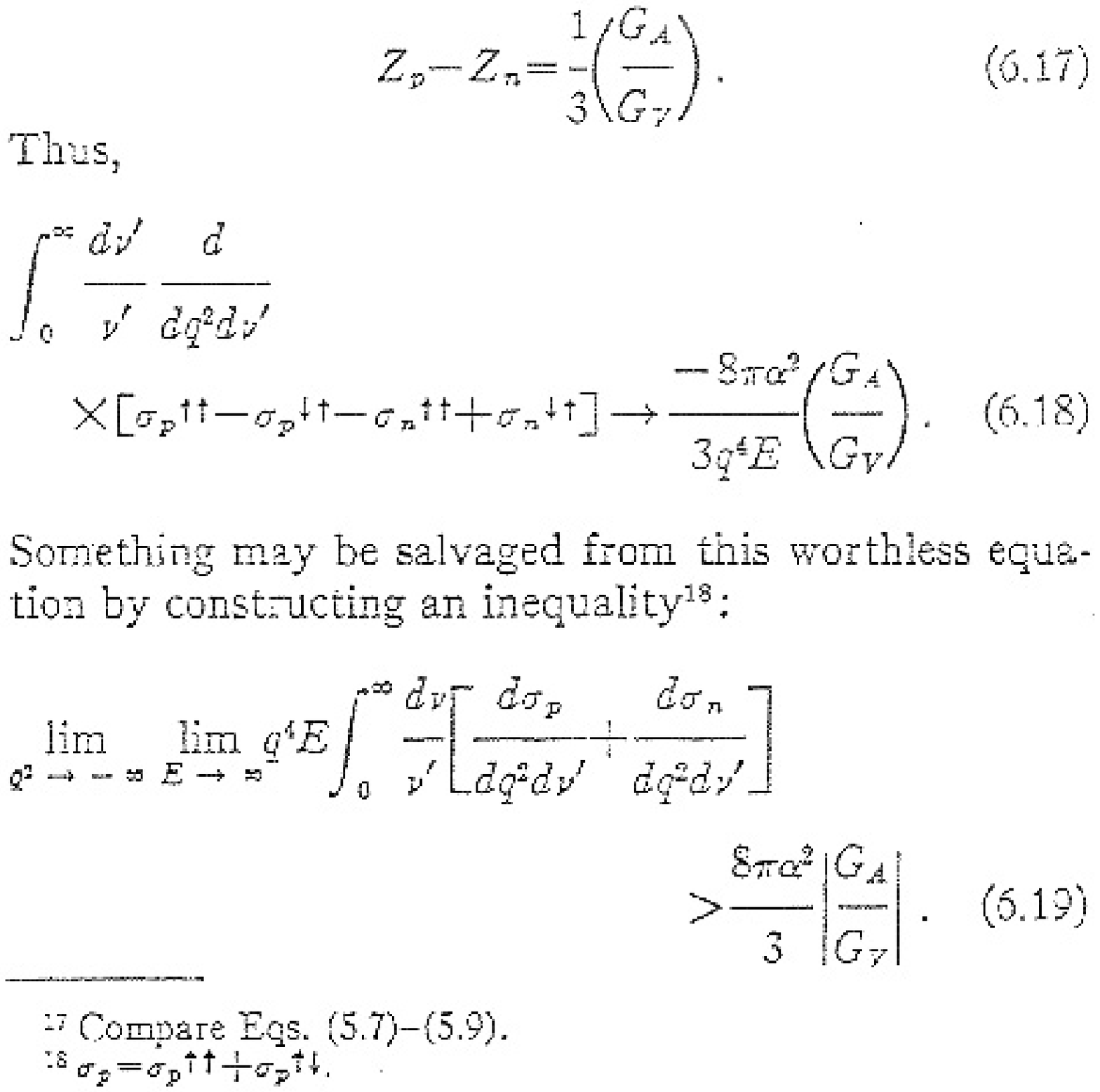 scaled 750}  
$$
\caption{From Bjorken's paper ``Applications of the
Chiral $U(6)\times U(6)$ Algebra\ldots.''}
\label{Bj1}
\end{figure}
Eq.~(6.18) is the famous Bjorken Sum Rule 
that has played an important role in my life.  As you can see, Bjorken 
didn't think much of his sum rule at the time.  Most likely he doubted 
that it could be tested experimentally.  So he changed the sign of 
these two negative terms and concluded that inelastic electron-nucleon 
scattering cross sections had to be greater than something that falls 
like a power of four-momentum transfer. Implicit in Eq.~(6.19) were the
ideas that led to scaling, to quarks,  and to the future.  So with this
equation, I think it is appropriate  to mark the end of the Archaic Era in
electron scattering.

\begin{figure}[htb]
$$
\BoxedEPSF{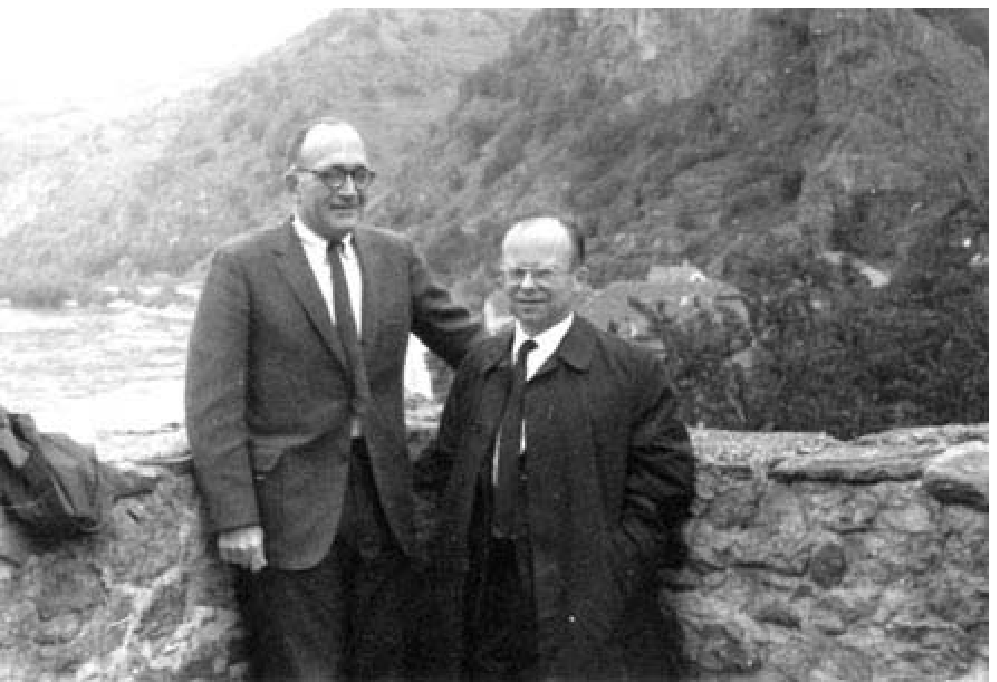 scaled 1400} 
$$
\caption{Sid and W.K.H. Panofsky, 1968.}
\label{Sid2}
\end{figure}

Returning to our unconventional time-keeping system, Figure~\ref{Sid2}
clocks in with Sid together with Pief Panofsky at the 1968 Vienna 
Conference, when the first results of the MIT-SLAC inelastic electron 
scattering experiments were announced in a parallel session by Jerry 
Friedman.

\section*{The Classic Era, 1966--1972}

\begin{figure}[htb]
$$
\BoxedEPSF{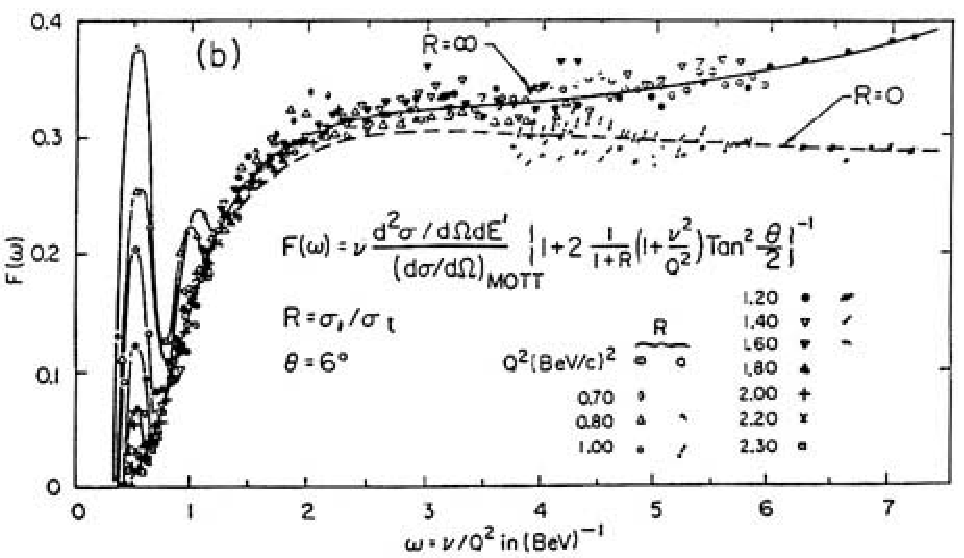 scaled 1400}  
$$
\caption{Approximate ``Bjorken'' scaling in 1968.}
\label{scaling}
\end{figure}

\begin{figure}[htb]
$$
\BoxedEPSF{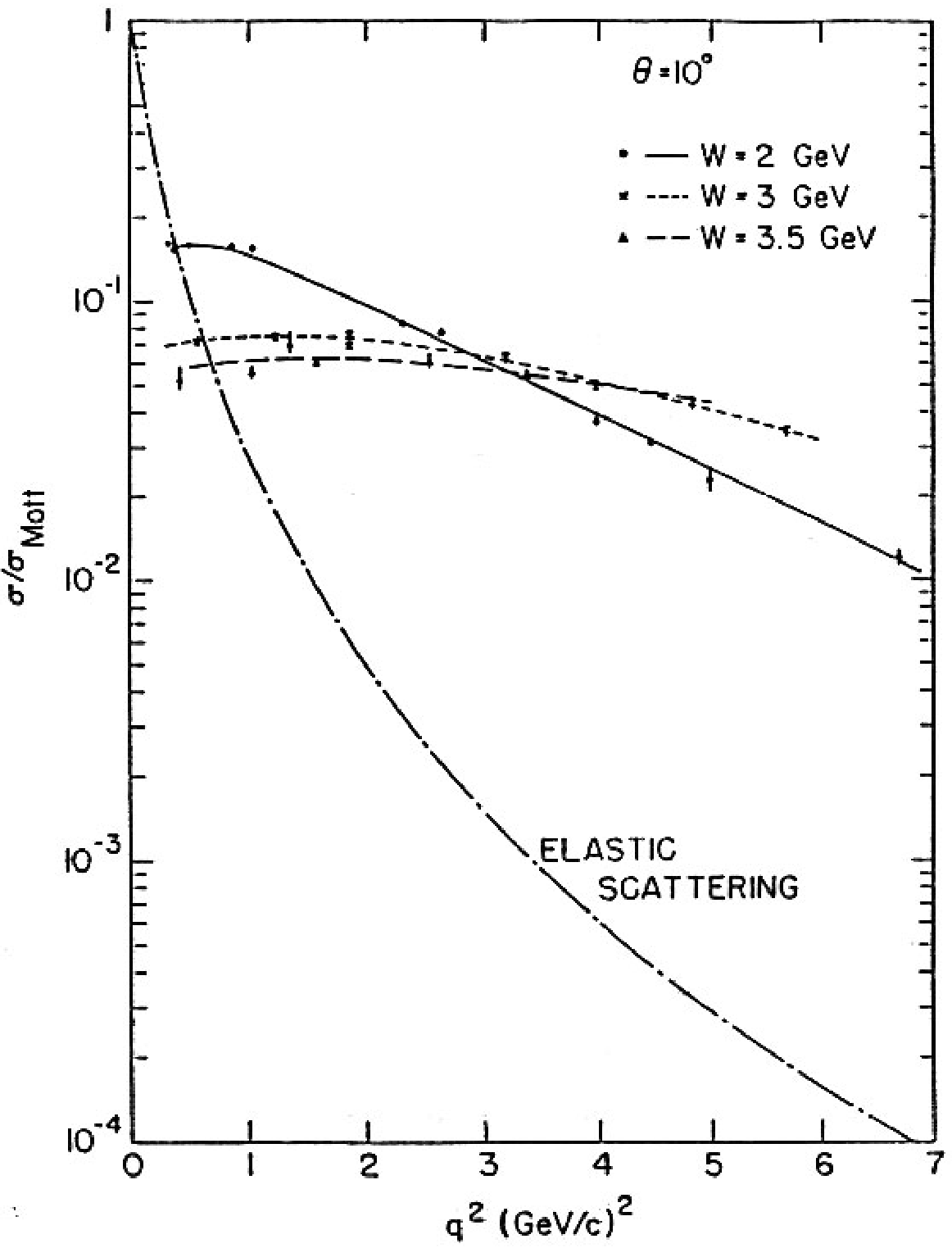 scaled 700}  
$$
\caption{Inelastic electron scattering data versus $Q^{2}$  
at fixed $W$.}
\label{qsquared}
\end{figure}

The 1968 Vienna Conference marked a turning point in this subject: the 
first public presentation of deep inelastic scattering data and the 
first direct evidence for pointlike constituents inside the nucleon.  
The MIT-SLAC group plotted their data in many ways.  The now famous 
plot, first suggested by Bjorken, and shown by Friedman in his 
parallel session talk~\cite{20}, is reproduced in Figure~\ref{scaling} and 
displays the approximate ``Bjorken'' scaling, which has been so 
important over the years.  At the time, Friedman found Figure ~\ref{qsquared} more
compelling.
It displays the data as a function of four momentum transfer ($Q^{2}$) 
in bins of fixed final-state mass, $W$.  Kinematic effects are removed 
by dividing the inelastic cross section by the cross section for 
scattering from a point charge.  For elastic scattering this isolates 
a combination of squares and products of the elastic form factors, 
which is shown for comparison.  Although the elastic cross section 
falls like a bomb compared to scattering from a point charge, the 
inelastic form factors stay large, indicating that there are 
pointlike objects inside the nucleon.

Even in these earliest results it was possible to see three essential 
features of deep inelastic scattering.  First was that the cross 
sections remained large at large momentum transfer, to which I have 
already alluded; second was that the structure functions scale; and 
third was that the structure functions are numerically small.  The 
last suggested that the pointlike constituents of the nucleon have 
fractional rather than integer charges.  It also played a role in 
understanding the momentum content of the proton, as Chris Llewellyn 
Smith will relate in his talk~\cite{21}.

You have all heard and read about this period at SLAC. Rather than 
attempt another general overview, I would like to describe what it 
looked like to me as a young theory graduate student newly arrived at 
SLAC in 1969.  Historians of science have pointed to the emergence of 
QCD as an example of a Kuhnian revolution in science~\cite{22}. 
Although I am skeptical about much that is written about modern 
particle physics by historians of science~\cite{23}, I believe 
they are correct in this interpretation of the developments that led to
QCD.  It is a fine example of a time when physicists were enthusiastically 
using ideas amalgamated from old and new models. They knew 
the ideas were contradictory, but they used them anyway and got
the right  answers.

To set the tone, consider Gell-Mann, who five years before had  said that
quarks were mathematical and hadrons were made up out of  each
other.  At the 1971 Coral Gables Conference  Gell-Mann said, ``Nature
reads books on free field theory''~\cite{25}.  So what  did he mean by
that?  At that time, he and Harald Fritzsch had  invented an ingenious
generalization of Bjorken's free-field  commutator algebra, where
commutators of operators at light-like  separation were given by free
field theory,
\begin{equation}
    \bigl[J_{\mu}^{a}(x),J_{\nu}^{b}(y)\bigr]\Bigr|_{x^{+}=y^{+}} 
    \Rightarrow\hbox {Free field 
    theory}.
\end{equation}
The Fritzsch--Gell-Mann algebra was accompanied by the ad hoc rule that
the matrix elements of  the operators that appear on the right hand side
of these commutators  should be taken parameters, and certainly not
from free field  theory.\footnote{Unless, of course, they are determined
by symmetries  of the theory.} This was a conservative approach, which
could not  be applied to processes other than deep inelastic scattering
(and $e^{+}e^{-}$ annihilation).

An alternative picture was the parton model of Feynman, Bjorken, and 
Paschos~\cite{26, 27}. Their's was a promiscuous approach -- in which 
one didn't know when to stop.  Basically, one used free field theory  in
momentum space to calculate cross sections, and used it wherever  one
wanted.  The rules were quite vague; theorists found themselves doing 
calculations that they could hardly believe they should be doing:
calculating strong interactions with free field theory -- and  getting the
right answer.\footnote{At least those with great physical  intuition, like
Bjorken and Feynman, got the right answer.}

Sid responded characteristically to this remarkable situation.  
Together with Tung Mow Yan and Don Levy, he undertook to construct a 
model field theory with a real Hamiltonian, that incorporated some 
minimal dynamical assumptions necessary to obtain Bjorken 
scaling~\cite{28}. They used infinite momentum frame methods, which 
soon afterward were canonized as light-cone field theory.  They 
introduced a transverse momentum cutoff -- the essential dynamical 
ingredient that guaranteed scaling.  We now recognize this as a crude 
but effective implementation of asymptotic freedom.  Armed with this 
model they could study other deep inelastic processes and learn where 
parton model results could be justified and where they could not.  
Some of the processes they studied had been studied before, but 
others, notably $pp\to \mu^{+}\mu^{-}X$ were new.  Now famous, the 
``Drell-Yan'' process is a mainstay of collider physics, and a 
subject of Tung Mow Yan's talk later this morning~\cite{29}.

\begin{figure}[htb]
$$
\BoxedEPSF{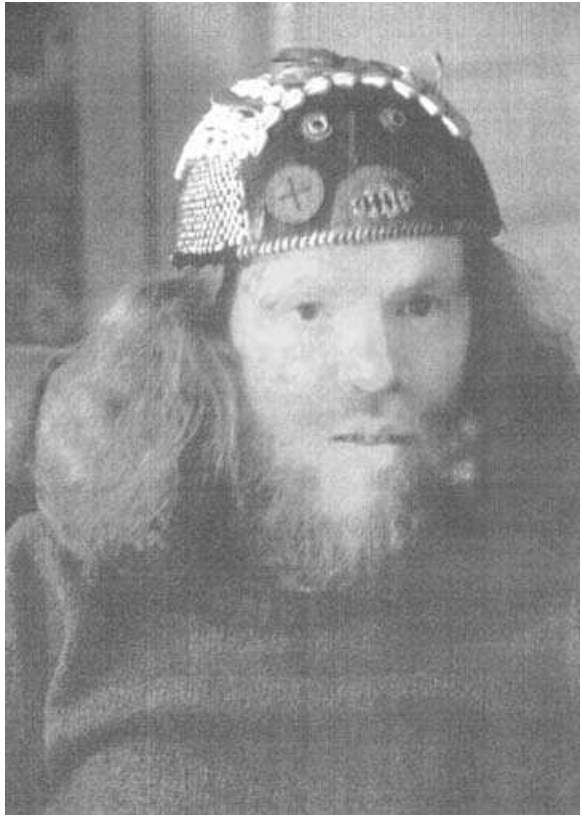}  
$$
\caption{John Ellis sometime after his arrival at SLAC.}
\label{Ellis}
\end{figure}

There was another way of approaching deep inelastic physics which was 
influential in the early 1970s, but has faded from memory somewhat 
more than the parton model or light-cone algebra.  It spread through 
the SLAC theory group after it arrived with the barbarian invasion I 
referred to earlier.  The approach is called ``canonical field 
theory'' and the barbarians brought it from England~\cite{30}.  
Figure~\ref{Ellis} shows you  one of the English invaders.  This particular  barbarian,
whom you may recognize, is John Ellis.  David Broadhurst, Frank Close, John Ellis, Chris
Llewellyn Smith, and Geoff  West arrived at SLAC in 1969 through 1971; Tony Hey, who
was at Caltech, was a frequent visitor.  Figuratively at
least, they mixed their intellectual genes  with the local population of
theorists, who included Bj,  Stan Brodsky, Sid of course, Fred Gilman, and
Haim Harari, and the  students, who included Mike Creutz, Inge Karliner,
Joe Kiskis, John  Kogut, Joel Primack, Dave Soper, and me.  

Canonical field theory was quantum field theory with a real 
Hamiltonian but without the infinities associated with radiative 
corrections.  We calculated deep inelastic processes either in 
coordinate or momentum space, whichever was more convenient; we 
checked that our results were preserved in models with nontrivial 
interactions; but we ignored the perturbative anomalies associated 
with renormalization (just as Bjorken had done in his 1966 paper).  
Those model field theories included quarks, color, vector gluons, soft 
chiral symmetry breaking -- all ingredients that we now 
associate with QCD. Canonical field theory amounted to QCD with zero 
anomalous dimensions, and the results, many of which Chris Llewellyn 
Smith will mention, are standard results that we now attribute to QCD.

A few summary remarks about this very creative period: First, I'd like 
to emphasize that the Bjorken-Johnson-Low method, not the operator 
product expansion, dominated thinking about short-distance physics at 
SLAC during this period.  Wilson's OPE did wonderful things for 
theoretical physics, but it was this other, equivalent stream of 
development that first led to scaling.  Second, the paper that did 
bring Wilson's methods (through the Callan Symanzik equations) to bear 
on deep inelastic scattering was a wonderful work by Norman Christ, 
Brosl Hasslacher, and Al Mueller that people were reading in 
1971~\cite{31}.

\begin{figure}[htb]
$$
\BoxedEPSF{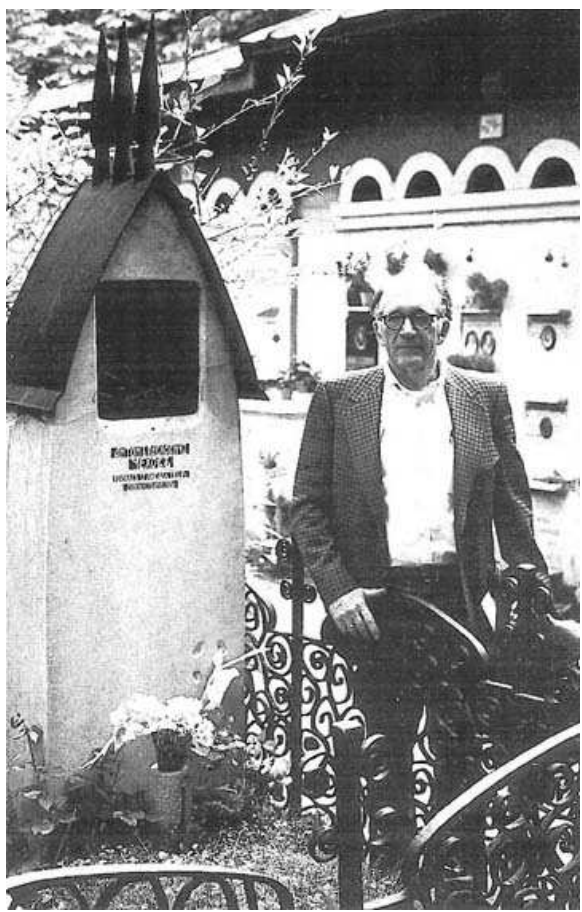}  
$$
\caption{Sid at the grave of Anton Chekhov in 1974.}
\label{Sid3}
\end{figure}

Thus, the classical era in electron-proton scattering had drawn to a 
close.  The major theoretical results now associated with QCD were in 
place in 1971 and 1972.  In the Kuhnian tradition, this occurred before 
QCD had been properly formulated and well before the discovery that 
QCD is asymptotically free.

So, onward to the Helenistic Period, when deep inelastic scattering 
spread to the far corners of the earth.  Returning to the Drell clock, 
there seem to be many pictures from this period, perhaps because 
photography had become more common.  I chose one in Figure~\ref{Sid3} 
because it is particularly meaningful to me -- I took it.  That's Sid 
on the right!  And on the left, more or less the same size and shape, 
is the tomb of Anton Pavlovich Chekhov.  Taken in the Novodevichy
Monastery, Moscow, in 1974, it shows Sid in one of SLAC's 
more distant colonies.

\section*{The Hellenistic Era, 1972--1980}

After the formulation of QCD came a period of great activity and 
discovery.  
In the early 1970s there was a great migration from SLAC to the 
provinces: Bjorn Wiik to Germany, Chris Llewellyn-Smith and John Ellis 
to CERN, Frank Close, Roger Cashmore, and others to England, 
many to Israel.  I myself migrated to the intellectual backwater of 
Cambridge, Massachusetts.  Deep inelastic scattering experiments were 
carried out with electron, muon and neutrino beams at DESY, CERN, and 
Fermilab, as well as at SLAC.

\begin{figure}[htb]
$$
\BoxedEPSF{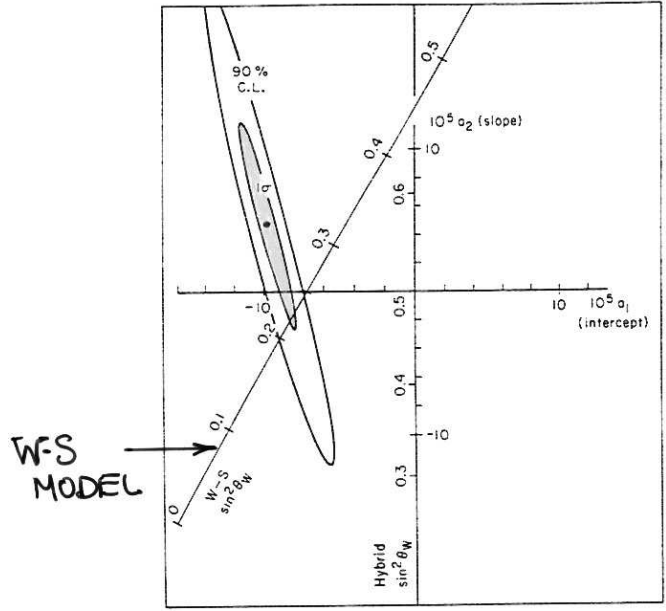 scaled 600}  
$$
\caption{Early data confirming the standard model prediction for $\vec
e d \to e'X$.}
\label{prescott}
\end{figure}

So much happened in the 1970s as the pieces of the Standard Model 
were put in place that it is impossible to do more than scratch the 
surface.  First I will mention a few of the developments that 
impressed me most, then point out some signs of things to come.  My 
first example is the decisive test of the Standard Model through 
parity violation in deep inelastic scattering of polarized electrons 
from unpolarized deuterium carried out at SLAC in the mid 1970s.  
This tour-de-force achieved by Charlie Prescott and collaborators 
confirmed the predicted $Z^{0}/\gamma$ interference and determined 
$\sin^{2}\theta_{W}$ to be near~1/4~\cite{32}. A summary of their 
data is shown in Figure~\ref{prescott}.  By this time the QCD-parton 
picture of deep inelastic scattering was so reliable that it formed 
the foundation for experimental analysis.

\begin{figure}[htb]
$$
\BoxedEPSF{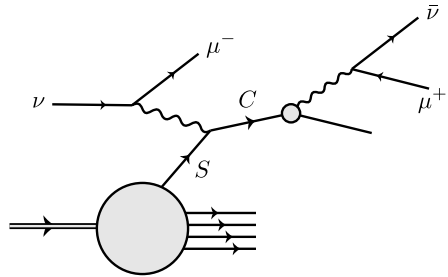 scaled 900}  
$$
\caption{Process by which opposite-sign dimuon production becomes a measure of the
strange quark distribution in the nucleon.}
\label{strange}
\end{figure}

Deep inelastic neutrino scattering experiments at CERN and Fermilab
were raising  puzzles and insights of their own.  Production of opposite-sign  dimuons
($\nu p \to \mu^{+}\mu^{-}X$),
 originally an anomaly,
was  understood as production and decay of charmed quarks, and finally 
turned into a measure of the strange quark distribution in the 
nucleon through the process shown in Figure~\ref{strange}. So dimuon 
production went from an anomaly to an experimentalist's tool in
short order.

\begin{figure}[htbp]
$$
\BoxedEPSF{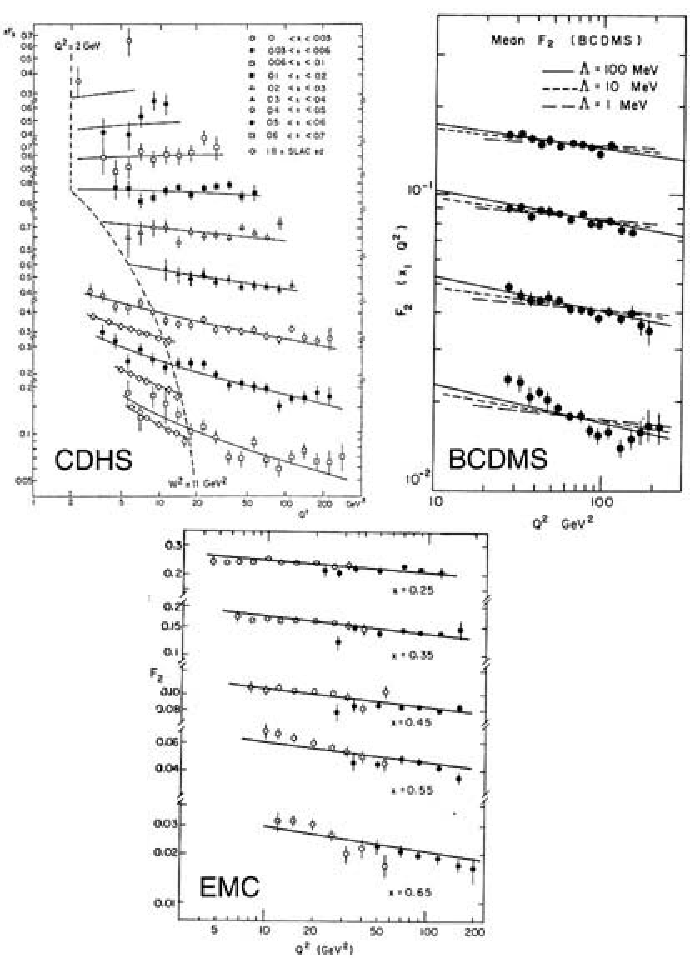 scaled 2000}  
$$
\caption{Logarithmic scaling violation in deep inelastic scattering.}
\label{ThreePreem}
\end{figure}

I can't survey this period without an homage to {\it the 
logarithms}.\footnote{Not to be confused with the ``Logarythms", 
one of MIT's better known choral groups.} Of course, the great {\it 
prediction\/} of QCD was not scaling, but scaling violation.  The 
theory predicted weak -- logarithmic -- dependence of the deep 
inelastic structure functions on squared four-momentum transfer.  In 
those days, every deep inelastic scattering collaboration was displaying 
its data versus $Q^{2}$ at fixed $x_{\rm Bj}$.  Data from that 
period from three of the preeminent experiments (EMC, CDHS, and 
BCDMS) are shown in Figure~\ref{ThreePreem}. The slopes of the fits to the
data are predicted by QCD, and the agreement between theory and 
experiment left little doubt that QCD is the correct theory of 
hadrons.  By 1980 QCD was established beyond a reasonable doubt; 
factorization theorems had quantified the parton model and ``canonical'' 
field theory; and we had learned what the transverse momentum cutoff 
in Drell, Levy, Yan really meant.

Now I would like to shift the focus away from the dominant ideas of 
the Hellenistic Era and onto three developments that were perhaps not 
so important at the time, but which played an important role in the 
future.  

\begin{figure}[htbp]
$$
\BoxedEPSF{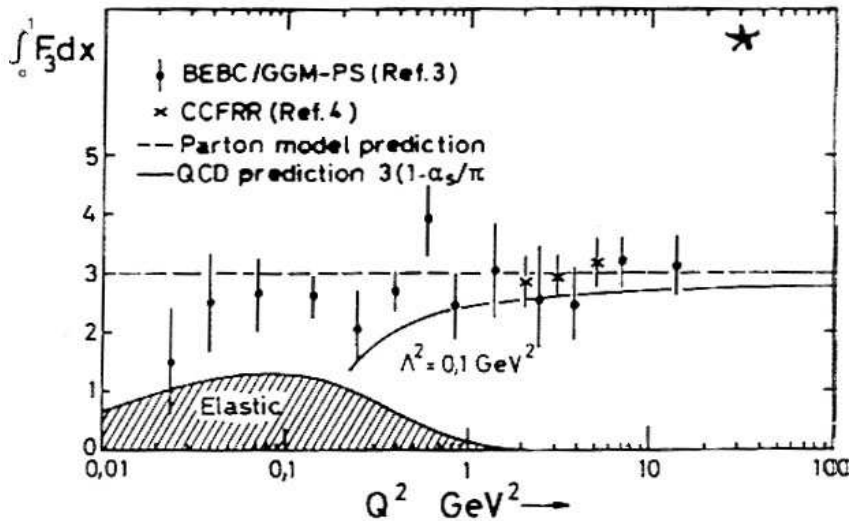 scaled 800}  
$$
\caption{From the Paris Conference, 1982.}
\label{GLS}
\end{figure}

The first was that scaling in electron scattering set in at very low 
$Q^{2}$.  This was first pointed out and analyzed by Elliott Bloom and 
Fred Gilman~\cite{34}. In retrospect, it was quite 
remarkable that scaling could have been discovered in the original 
MIT-SLAC experiments which were dominated by $Q^{2}\approx 1$ 
GeV$^{2}$, when scaling is supposed to be an asymptotic phenomenon.  A
couple of simple examples illustrate the point: First  consider the
asymptotic equipartition between quark and gluon  ``momentum'' --
actually $P^{+}$ -- in any hadron.  A famous  prediction of QCD is that
momentum should be shared roughly equally  between quarks and
gluons at asymptotically large $\log  Q^{2}$.\footnote{The prediction is
exact, $3n_{f}/16$, where
$n_{f}$  is the number of ``active'' quark flavors -- four in the early
days,  now five.} This prediction presumes that terms suppressed only
by logs  of $Q^{2}$ are ignorable, so all moments (in $x_{\rm Bj}$) of the 
structure functions except the lowest are negligible -- the structure 
functions are $\delta$-functions at $x_{\rm Bj}=0$.  One would therefore 
expect to see equipartition only at the highest $Q^{2}$ available, 
perhaps at HERA in the 1990s.  Instead, it works pretty well at 1 
GeV, excellently at 10 GeV. Why is asymptopia so close?  A second 
impressive example comes from the Gross-Llewellyn Smith Sum Rule named 
in part after the next speaker~\cite{35}. Figure~\ref{GLS} is taken
from the 1982 Paris Conference presentation, after an analysis by Bill 
Scott~\cite{36}. At large $Q^{2}$ the integral of the chiral-odd 
neutrino structure function, $F_{3}$ should asymptote to three.  As the 
figure shows, it equals to three at $Q^{2}$'s well below 1~GeV, where 
much of the contribution comes from the elastic peak.  The corrections 
to this sum rule can be calculated up to an overall constant that 
measures a correlation between quarks and gluons in the proton.  
Apparently that constant is negligibly small.

These examples give us a modern restatement of the old puzzle that 
first surfaced with the measurement of the elastic form factor: Why is 
the proton so soft?  The proton is very composite even though it is 
made of pointlike objects.  The distributions of the nucleon's 
constituents are only very weakly correlated.  Once confinement is 
taken care of (by constructing color singlets) we see little evidence 
for non-perturbative interactions between quarks.  The proton looks 
very much like a bag of confined but otherwise free quarks, an idea 
that Sid and I pursued intensely on during this period~\cite{37,38}.

\begin{figure}[htbp]
$$
\BoxedEPSF{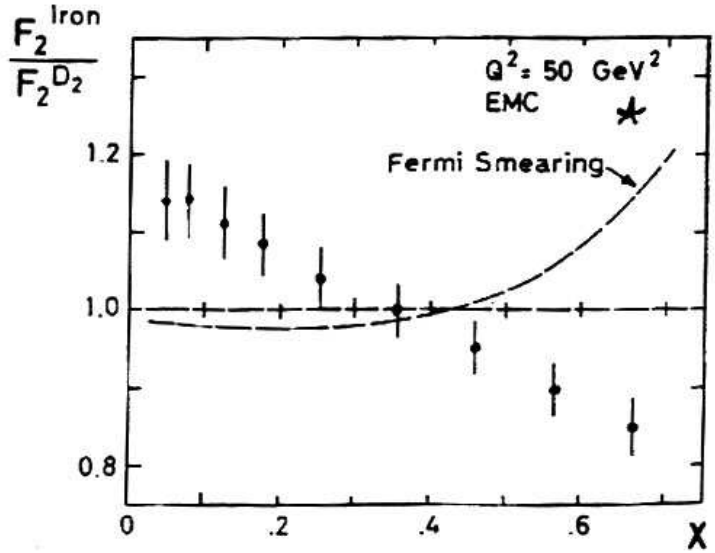 scaled 850}  
$$
\caption{First results on the ratio of nuclear to nucleon structure
functions.}
\label{emcdata}
\end{figure}

Another hint at the future came with the first precision measurements 
of deep inelastic scattering off nuclei~\cite{39}. The European Muon 
Collaboration at CERN scattered muons from iron and deuterium and 
compared the structure functions.  They didn't expect to see much -- 
principally the effects of Fermi motion.  In Figure \ref{emcdata}
Fermi motion is the dashed curve.  Instead, they found a strong 
$x_{\rm Bj}$ dependence with the opposite slope from Fermi motion.  The 
data were first shown at the 1982 Paris Conference, where they 
attracted very little attention.  The elaboration of this subject 
properly belongs in the next section of my talk, where I will return 
to it.

Also at this time, SLAC ran a small program in deep inelastic 
scattering of polarized electrons from polarized protons.  Polarized 
beam was required for the parity violation program, while the 
polarized target technology was new.  The object was to measure the 
helicity distribution of quarks in the polarized target.  Vernon 
Hughes and his group from Yale joined Dave Coward and others at SLAC 
in this enterprise.  
\begin{figure}[htbp]
$$
\BoxedEPSF{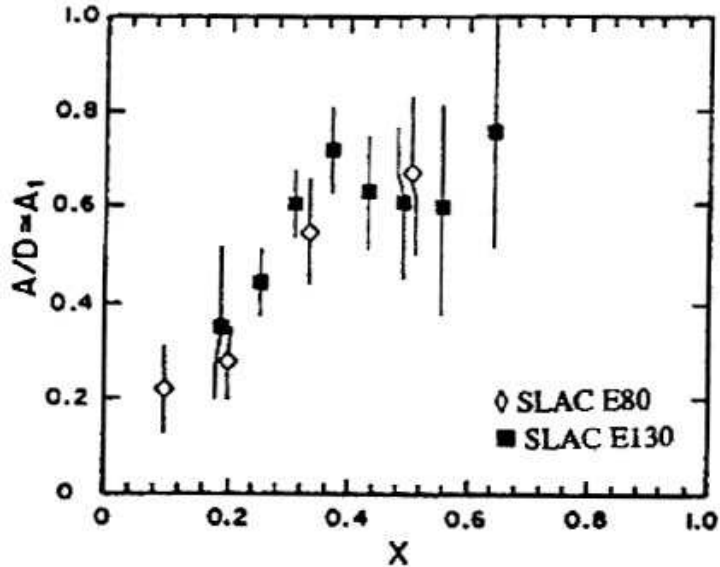 scaled 1000}  
$$
\caption{Early SLAC data on the deep inelastic spin asymmetry.}
\label{E130}
\end{figure}    
 Data from their two experiments, E80 and E130, are shown in
Fig.~\ref{E130}~\cite{40}. This data attracted rather little interest,
although its  importance as a probe of the spin content of the proton was
``well  known among a small group of people''.

After E130, the SLAC program turned in other directions.  Polarized 
scattering at SLAC was focused on parity violation; future development 
focused on $e^{+}e^{-}$ collider physics.  Deep inelastic scattering 
was no longer SLAC's highest priority.  Hughes's request for 
extensions of the polarized target program were turned down by the 
SLAC PAC. So, perhaps this is a good point to mark the end of the 
robust expansion of deep inelastic scattering physics -- the end of 
the Hellenistic Era.  QCD was supreme; deep inelastic scattering was 
routine; and Vernon Hughes was a prophet cast out in the desert.

Well, next is the Renaissance.  It seems appropriate to clock in with 
Sid enjoying a conversation with his granddaughter Cornelia.  The date 
is approximately 1988, just when new energy was appearing in the 
electron-nucleon scattering community.
\begin{figure}[htbp]
$$
\BoxedEPSF{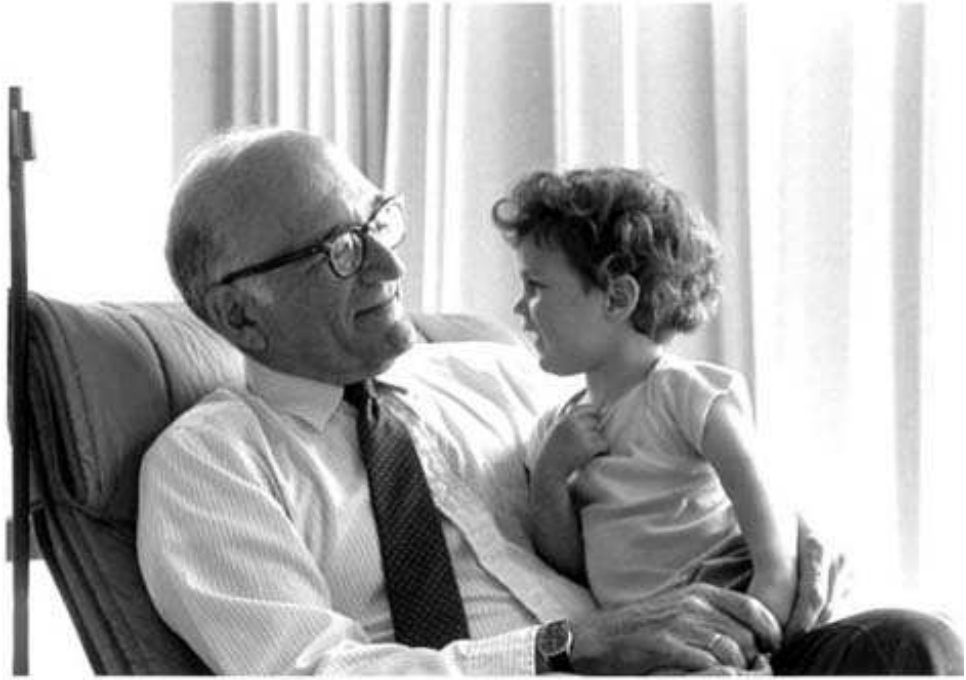 scaled 1200}  
$$
\caption{Sid with his granddaughter Cornelia in 1988.}
\label{Cornelia}
\end{figure}    

\section*{Renaissance,  1983--Today}

Interest in deep inelastic scattering and related phenomena (including 
even form factors) has been reborn in recent years because they provide a 
unique probe of the mysteries of confinement.

People no longer seriously doubt the basic validity of QCD, although a 
few rogues remain unconvinced.  Considerable interest in QCD remains 
because we must understand the quark and gluon content of beams, 
targets and final state fragments in order to identify novel phenomena 
beyond the Standard Model.  This effort is very important in the 
search for physics beyond the Standard Model, but as a sort of ``QCD 
engineering'', it is perhaps not as intellectually exciting as the early 
days of deep inelastic scattering.

Although QCD is very familiar, and models abound, confinement is still 
not understood.  QCD is very well understood at short distances, where 
perturbative methods apply.  However, at long distances, where hadrons 
form, our theoretical tools are few.  This seems like an 
inadequate state of affairs for such a beautiful problem.  After 
all, the Lagrangian of QCD is extraordinarily simple,
\begin{equation}
    {\cal L} = -{\textstyle\frac{1}{4}} {\rm Tr }\  F^{2} + \bar q (i\thruD{D} - m )q\ .
\end{equation}
I have suppressed a few sums over degrees of freedom, but this 
Lagrangian can be written on a postage stamp, not to mention a t-shirt. 

We need theoretical tools to understand confinement.  To develop them 
we need probes of hadrons that give precise, interpretable 
information.  In the old days Sid used to say (perhaps this is even a 
direct quote), ``Leptons make great probes of hadrons because leptons 
are pointlike.''  The Renaissance that deep inelastic physics has 
enjoyed over the past 15 years came from the realization that we now 
have a new version of that old slogan: ``Quarks and gluons make great 
probes of hadrons in hard processes, because at large momentum 
transfer, quarks and gluons are pointlike.''  We can use perturbative 
QCD, which we understand at short distances, to probe the aspects of 
confinement at long distances that so far defy understanding.  This 
helps organize the most interesting developments in electron 
scattering and related subjects in this fourth period.  Let me close 
my talk by giving some examples.

The first concerns how quarks are distributed in nuclei.  Someone once  said,
anonymously, that ``Looking for quarks in the nucleus is like  looking for the
Mafia in Sicily.  Everyone knows they're there, but  it's hard to find the
evidence.''  The first sign that the quark  distribution in a nucleus differs
significantly from the distribution  in isolated nucleons came from the
European Muon Collaboration data  shown in Figure~\ref{emcdata}.  

\begin{figure}[htbp]
$$
\BoxedEPSF{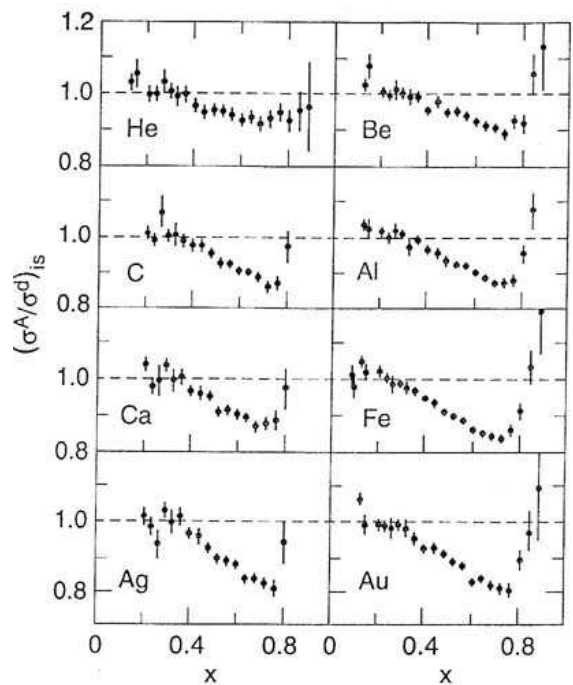 scaled 800}  
$$
\caption{Electron scattering from a variety 
of nuclei: SLAC E139.}
\label{Bodek}
\end{figure}

\begin{figure}[hb]
$$
\BoxedEPSF{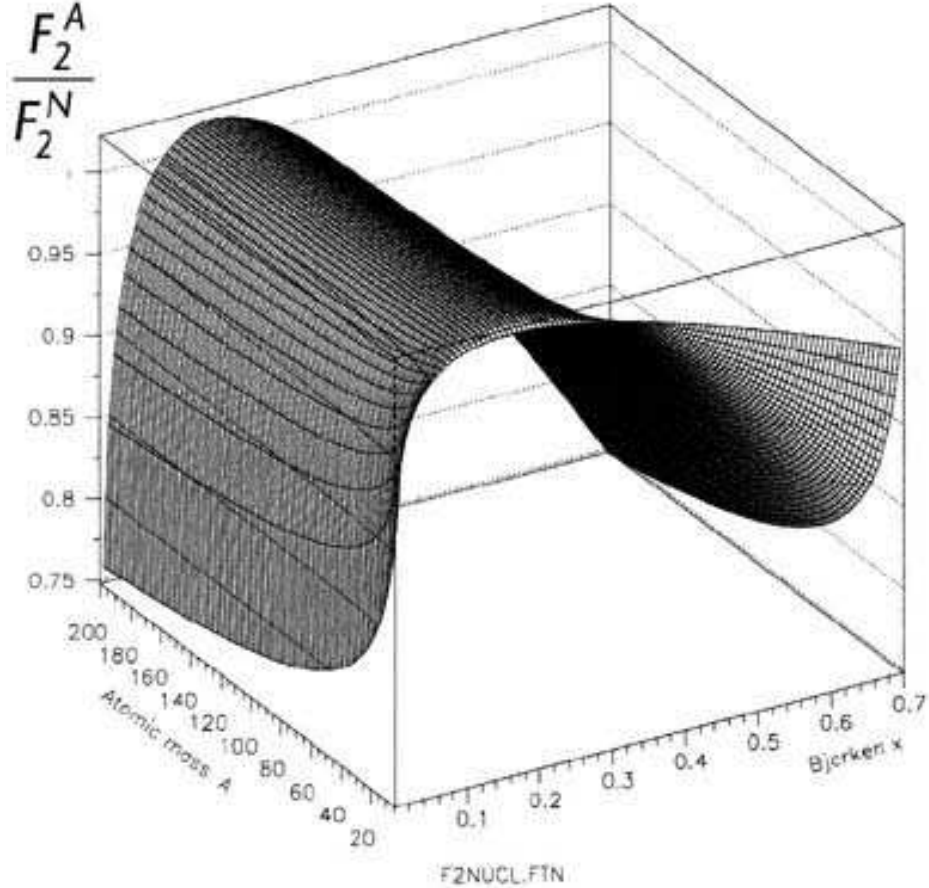 scaled 1400}  
$$
\caption{A fit to the world data on the ratio of nuclear to nucleon structure
functions.}
\label{global}
\end{figure}

As you
know from freshman world history,  during
{\it the\/} Renaissance, people became interested in  archaeology.  They
went out and dug around to see what happened before  them.  In a
marvelous example of particle physics archaeology, Ari  Bodek of the
University of Rochester and his collaborators did exactly this in order to 
check the EMC results. 
He exhumed the old SLAC deep inelastic scattering data.  The targets 
were hydrogen and deuterium, which, of necessity had been contained in 
metallic vessels.  In order to subtract the background scattering on 
the vessels, ``target empty'' data had been taken off the empty 
vessels which were aluminum or iron.  So, without applying to SLAC's 
program advisory committee, Bodek was able to get SLAC data on deep 
inelastic electron scattering off nuclei~\cite{42}.  The data proved so 
interesting that he and his collaborators were then able to launch a 
new program at SLAC to measure electron scattering from a variety 
of nuclei.  The data obtained this way are shown in Figure \ref{Bodek}.
It may be difficult to extract the trends from these 
separate curves, so instead consider Figure~\ref{global}, the global
fit to nuclear data recently done by Gueorgui Smirnov~\cite{43}, 
which shows clearly the power of deep inelastic scattering.  The ratio 
of structure functions of nuclei to deuterium is shown as a function 
of $x_{\rm Bj}$ and A, atomic mass, from $A=20$ to $A=200$.  There are 
four distinct domains.  At very low $x_{\rm Bj}$ there is shadowing -- quarks do 
not see the whole nucleus.  Beyond the range of $x_{\rm Bj}$ shown, the 
cross sections shoot up, because of Fermi motion: the nuclear 
structure functions don't have to vanish at $x_{\rm Bj}=1$ but the proton 
and neutron structure functions do, so the ratio must go to infinity.  
The intermediate region $0.3< x_{\rm Bj} < 0.7$ held the greatest 
surprise.  In this region there is a systematic suppression of the 
nuclear structure function relative to the free nucleon.  The final 
domain, where the ratio exceeds one at moderately low $x_{\rm Bj}$, is 
required by sum rules to compensate the depletion at larger $x_{\rm Bj}$.

The depletion at large $x_{\rm Bj}$ and the corresponding enhancement 
below show that quarks are systematically 
shifted from large $x_{\rm Bj}$ to small $x_{\rm Bj}$ by the presence of the 
nuclear medium.  As first pointed out by Boris Ioffe in the late 
1960s~\cite{44}, $x_{\rm Bj}$ is conjugate, in the uncertainty principle sense, to a 
correlation length along the light-cone.  Thus a shift to lower 
$x_{\rm Bj}$ means that the quark ``mean free path'' along the light-cone 
has increased in nuclei.  They propagate further along the light-cone 
between the absorption and re-emission of the deeply virtual photon 
exchanged in deep inelastic scattering.  One can say, quite 
independent of the specific dynamical model, that quarks are less 
severely confined in nuclei.  Nearly as many mechanisms have been 
proposed as papers written on this subject.  Perhaps the microscopic 
origin lies in an effective medium dependent change in
$\Lambda_{QCD}$, or perhaps in quark exchange between nucleons.  Other 
papers study meson exchange.  All these are only models, but the basic 
fact remains: quarks are shifted to lower $x_{\rm Bj}$ because they are 
partially deconfined in the nuclear medium.

Another beautiful and exciting example of the use of deep inelastic 
phenomena to explore confinement concerns the distribution of angular 
momentum within the nucleon.  First studied by Vernon Hughes and 
collaborators during the Hellenistic Era, the Renaissance began with 
measurements of deep inelastic scattering of polarized muons from 
polarized nucleons by the European Muon Collaboration at 
CERN~\cite{45}. Not 
surprisingly, Hughes played a major role, first as collaborator and 
then as co-spokesman for experiments that mapped out the quark spin 
distributions in the proton and neutron over the period 1986--1996.  
The EMC discovered, and later experiments at CERN, SLAC, and HERA
confirmed,  that only about 30\% of the spin of the proton is carried by the
spin  of the quarks.  This was quite a surprise, since expectations based 
principally on the hypothesis that polarized $s$-quarks are not 
important, suggested something closer to 60--65\%.  This puzzle, known 
affectionately as the violation Ellis-*****~Sum Rule, will be explored 
in some depth by Chris Llewellyn Smith in the next talk~\cite{46}. Just 
\begin{figure}[hb]
$$
\BoxedEPSF{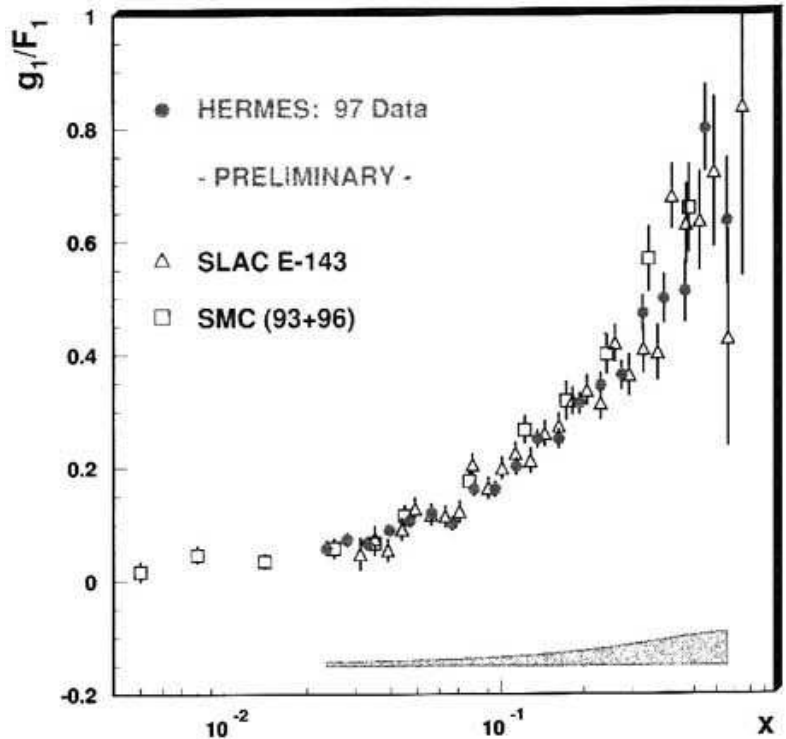}  
$$
\caption{The spin asymmetry measured at  HERA (by the Hermes 
Collaboration) together with data from SMC and SLAC.}
\label{Hermes}
\end{figure}       
to give you an idea of how far this field has progressed, 
Figure~\ref{Hermes} shows the spin asymmetry measured at
 HERA (by the Hermes 
Collaboration) together with data from SMC and SLAC~\cite{47}. 

The study of spin-dependent deep inelastic phenomena has moved beyond 
the initial excitement engendered by the quark spin measurement.  It 
has blossomed in the past decade and is one of the richest areas in 
QCD research at this time~\cite{47.1}.  Experiments underway at CERN and at HERA 
hope to determine the gluon spin distribution in the nucleon.  There 
is a new program at the Brookhaven Relativistic Heavy Ion Collider 
(RHIC) aimed at unravelling the quark and gluon spin structure of the 
nucleon~\cite{48}.  One of the prettiest ideas is to use $\vec p p \to W^{\pm} 
X$, known as polarized-Drell Yan, to measure the spin distribution of 
$u$ and $d$ quarks in the nucleon separately~\cite{49}.  The simple dynamics of 
the Drell Yan process and the parity violation in $W^{\pm}$ production 
make this possible.  Finally, a whole new understanding of the deep 
inelastic physics of transverse spin has emerged from intense study of 
polarization phenomena.  We now understand, for example, that a 
complete understanding of the quark spin in the nucleon in the deep 
inelastic limit requires measurement of a third quark polarization 
distribution, known as the ``transversity'' distribution~\cite{50}.  It 
describes the distribution of transversely polarized quarks in a 
nucleon polarized transverse to its direction of motion.  Plans are 
underfoot to measure transversity at HERMES, at HERA, and at RHIC, 
where one possibility is $\vec p_{\perp} \vec p_{\perp} \to \mu 
\bar\mu X$, known appropriately as transverse-Drell, transverse-Yan.

My final example harkens back to the earliest days of this culture, 
back to the 1950s.  It turns out that there are still very 
interesting things to be learned from elastic electron-nucleon 
scattering.  The electromagnetic current, which has been so well 
studied in elastic scattering, is a pure SU$(3)_{f}$ octet.  By 
exploiting flavor symmetries one can separate two flavor structures: 
the isovector, $u-d$, and the hypercharge, $u+d-2s$.  Because it has 
no singlet component, no information on $u+d+s$ can be obtained, so 
there is no way to separate the $u$, $d$, and $s$ contributions to any 
matrix element.  The $Z^{0}$ offers a new flavor coupling to the 
nucleon proportional to weak isospin, which samples $u-d-s$ in the 
light quark sector~\cite{51}.  The $c$, $b$, and even $t$ quarks also 
contribute, but their effects can be handled in the heavy quark 
approximation by the methods of perturbative QCD. Since the matrix 
elements of $u-d$ are already known, the $Z^{0}$ effectively probes 
strange quark matrix elements in the nucleon.  The hitch is that 
$Z^{0}$-nucleon couplings can only be measured in elastic neutrino 
scattering or in parity violating elastic electron nucleon scattering 
-- both extremely difficult experiments.  For example, the parity 
violating asymmetries in elastic $eN$ scattering are of order 
$10^{-7}$.  The rewards are great, however, since one can measure the 
strangeness analogues of the axial charge ($\bar s 
\vec\gamma\gamma_{5} s$), magnetic moment ($\frac{1}{2}s^{\dagger} 
\vec r \times \vec \alpha s$), and ``charge'' radius ($s^{\dagger} 
r^{2} s$).  The first would provide an independent confirmation 
of the quark spin fraction measurements without the difficulties of 
extrapolation to $x_{\rm Bj}=0$ and use of exact $SU(3)_{f}$ symmetry.  
The latter two are otherwise unknown measures of strangeness in the 
nucleon.  Major experiments, which are the direct descendents of the 
program begun nearly half a century ago at Stanford by Robert 
Hofstadter, are underway to measure these quantities at low-energy 
electron machines.  The first accurate measurements of the strangeness 
magnetic moment have been announced by the SAMPLE collaboration 
working at Bates near Boston~\cite{52}.  Further experiments are being mounted 
at Jefferson Lab in Virginia.

I have reached the end of my survey.  In retrospect, I think Sid is a 
very lucky man: to have watched not only the acceptance but also the 
apotheosis of his ideas.  Electron scattering has gone from a relative 
backwater to the principal paradigm for understanding in particle 
physics.  All those battles for funding that Sid helped to wage now 
appear vindicated.  As for pedagogy, Mike Peskin has already spoken of 
Sid's wonderful book with Bj.  I would only add that when a new book 
on quantum field theory appears -- by Weinberg, or Brown, or Peskin 
and Schroeder -- the question everyone asks is ``Will it be the next 
Bjorken and Drell?''

Personally, I would like to add that I can hardly imagine physics at 
SLAC without the memory of Sid's gruff voice echoing down the 
corridors of the third floor to herd students toward the Green Room 
for the Friday lunchtime student seminar.  Sid, as many of you know, 
was the only person with a PhD allowed in the room, because he promised 
to ask all the stupid questions that students were too shy to ask.  
And he did.  Thank you, Sid for all that, and thank you all.

\section*{Acknowledgments}

I would like to thank Jim Bjorken, Jerry Friedman, Henry Kendall, and Chris Llewellyn
Smith for help with the background of this talk. Thanks  also to Mike Peskin and the
conference organizers for the invitation to speak and for their support, and especially to
Harriet Drell for providing wonderful pictures of Sid. 
\bigskip

\def\Journal#1#2#3#4{{#1} {\bf #2}, #3 (#4)}
\def\add#1#2#3{{\bf #1}, #2 (#3)}

\def\NPB{{\em Nucl. Phys.} B}
\def\PLB{{\em Phys. Lett.}  B}
\def\PRL{{\em Phys. Rev. Lett.}}
\def\PRD{{\em Phys. Rev.} D}
\def\PR{{\em Phys. Rev.}}
\def\ZPC{{\em Z. Phys.} C}
\def\SJNP{{\em Sov. J. Nucl. Phys.}}
\def\AnnP{{\em Ann. Phys.}}
\def\JETPL{{\em JETP Lett.}}
\def\LMP{{\em Lett. Math. Phys.}}
\def\CMP{{\em Comm. Math. Phys.}}
\def\PTP{{\em Prog. Theor. Phys.}}


\end{document}